\documentclass[american,aps,pra,reprint,superscriptaddress, longbibliography]{revtex4-1}
\usepackage[T1]{fontenc}
\usepackage[latin9]{inputenc}
\usepackage{color}
\usepackage{babel}
\usepackage{amsthm}
\usepackage{amsmath}
\usepackage{amssymb}
\usepackage{graphicx}
\usepackage[unicode=true,pdfusetitle, bookmarks=true,bookmarksnumbered=false,bookmarksopen=false,  breaklinks=false,pdfborder={0 0 0},backref=false,colorlinks=false] {hyperref}
\hypersetup{ colorlinks,linkcolor=myurlcolor,citecolor=myurlcolor,urlcolor=myurlcolor}

\makeatletter
\@ifundefined{textcolor}{}
{%
 \definecolor{BLACK}{gray}{0}
 \definecolor{WHITE}{gray}{1}
 \definecolor{RED}{rgb}{1,0,0}
 \definecolor{GREEN}{rgb}{0,1,0}
 \definecolor{BLUE}{rgb}{0,0,1}
 \definecolor{CYAN}{cmyk}{1,0,0,0}
 \definecolor{MAGENTA}{cmyk}{0,1,0,0}
 \definecolor{YELLOW}{cmyk}{0,0,1,0}
 }
\theoremstyle{plain}

\theoremstyle{plain}

\ifx\proof\undefined
\newenvironment{proof}[1][\protect\proofname]{\par
\normalfont\topsep6\p@\@plus6\p@\relax
\trivlist
\itemindent\parindent
\item[\hskip\labelsep
\scshape
#1]\ignorespaces
}{%
\endtrivlist\@endpefalse
}
\providecommand{\proofname}{Proof}
\fi
\theoremstyle{plain}

\providecommand{\lemmaname}{Lemma}
\providecommand{\definitionname}{Definition}
\providecommand{\propositionname}{Proposition}

\usepackage{babel}
\usepackage{txfonts}
\usepackage{colortbl}\definecolor{myurlcolor}{rgb}{0,0,0.7}
\newcommand{\tr}{{\operatorname{Tr\,}}}

\newcommand{\ketbra}[2]{|#1\rangle\!\langle#2|}
\def\ket#1{| #1 \rangle}
\def\bra#1{\langle  #1 |}
\def\braket#1{\langle  #1 \rangle}
\def\proj#1{| #1 \rangle\!\langle #1 |}

\DeclareMathOperator{\Tr}{\mathrm{Tr}}

\newcommand{\haH}

\definecolor{orange}{RGB}{255,127,0}

\begin{document}
\title{ Quantifying Superpositions Between Quantum Evolutions} 
\author{Manabendra N. Bera}
\email{mnbera@gmail.com}
\affiliation{ICFO -- Institut de Ci\`encies Fot\`oniques, The Barcelona Institute of Science and Technology, ES-08860 Castelldefels, Spain}
\affiliation{Max-Planck-Institut f\"ur Quantenoptik, D-85748 Garching, Germany}

\begin{abstract}
Quantum mechanics allows coherent superposition between different states of matter. This quality is responsible for major non-classical phenomena that occur in quantum systems. Beyond states, coherent superpositions are also possible between quantum evolutions. We characterize such superpositions here. 
A resource theoretic framework is developed to quantify superposition present in an arbitrary quantum evolution.
In addition to characterization, the framework considers superposition as a quantum resource. This resource can be exploited to perform certain quantum tasks that are otherwise impossible. We identify maximally resourceful evolutions and  demonstrate how these could enable one to implement arbitrary quantum operations and super-operations. We also discuss the roles of superposition to exhibit non-classical behaviors present in evolutions, for example, a-causality, temporal Bell correlations, and indefinite temporal and causal orders. 
\end{abstract}

\maketitle


\section{Introduction}
Possibility of coherent superpositions between different states of matter plays a central role in quantum mechanics. Much of the bizarre, non-classical features present in the quantum world can be attributed to quantum superpositions, for example wave-particle duality \cite{Cohen91}, quantum randomness and uncertainty associated with measurement \cite{Wheeler14, BeraPhilo16},  exotic quantum phases of matter \cite{Sachdev00, Amico08},  nontrivial correlations between quantum objects such as quantum  nonlocality \cite{Bell64, Brunner14}, contextuality \cite{Kochen67} and entanglement \cite{Horodecki09}, to mention a few. A systematic understanding of superposition and its role in various quantum phenomena are therefore essential not only to decode fundamentals of quantum physics but also for their applications in quantum technologies.  

Though it is as old as quantum mechanics, quantitative understanding of coherent superposition between quantum states has been done only recently \cite{Aberg06, Baumgratz14, Theurer17, Streltsov17}, in terms of resource theory -- an operational framework developed to characterize useful resources in quantum information theory. This information theoretic approach towards superposition brings alternative views and understanding towards a wide range of phenomena and applications, for examples wave-particle duality \cite{Bera15a, Bagan16}, roles in creating entanglement \cite{Streltsov15, Chitambar16a} and quantum correlations \cite{Ma16}, speed of quantum evolution and metrology \cite{Marvian16}, symmetries \cite{Marvian14}, quantum thermodynamics \cite{Lostaglio15, Lostaglio15a}, and certain information theoretic tasks \cite{Chitambar16, Streltsov16, Streltsov17a}, etc.

Analogous to correlations, like non-locality and entanglement, between space-like separated quantum objects, there also exists nontrivial correlations between two time-like separated quantum events. Even though each event respects certain local causal orders, their joint quantum events may not possess a definite causal order \cite{Oreshkov12, Brukner14, Branciard16, Rubino17}. There are also nontrivial correlations between different temporal directions of quantum operations, which leads to indefinite temporal order \cite{Procopio15, Zych17}. These nontrivial correlations between quantum events are shown to have important implications in understanding quantum theory of gravity \cite{Hardy05}, and many quantum information and computation tasks \cite{Chiribella12, Chiribella13, Araujo14, Feix15, Guerin16}. Like in quantum states, these nontrivial phenomena can be attributed to certain types \emph{coherent superpositions between quantum evolutions} too. The superposition in quantum evolutions, for the first time, has been noted in \cite{Aharonov90}. There, it has been shown that a time evolution of a quantum state can be composed by superposing two different evolutions. Later, information theoretic aspects of quantum evolutions has been explored in \cite{Oppenheim04}. However a systematic characterization of such superpositions is still pending.

There are two kinds of physical processes that a quantum object goes through, which are discussed in Appendix \ref{App:QuntEvol}. The first kind is quantum \emph{operations} \cite{Nielsen00} that act on quantum states and transform initial states to final ones. These operations are mathematically expressed by completely positive trace preserving maps, or more generally, completely positive trace non-increasing maps. The second kind is quantum \emph{super-operations} \cite{Chiribella08}, which operate on initial quantum operations and transform them to the final ones. Both of these processes can be implemented deterministically and probabilistically (or selectively), by means of quantum circuits (see Figs. \ref{fig:QuantOp} and \ref{fig:QuantSupOp}).

In this work we systematically characterize superpositions between quantum evolutions and, to do so, we introduce a resource theoretic framework. 
Indeed, superposition is a basis dependent quality. An evolution, that is in coherent superposition of certain evolution bases, can be made completely superposition-free by suitably choosing a new set of evolution bases. In many situations, these choices of evolution bases appear very naturally and often depend on underlying physical constraints. In case of quantum state, these can be constrained by system Hamiltonians or reference frames. While for quantum evolutions, these may depend on the availability of implementable quantum gates or operations in quantum circuits, or accessible driving Hamiltonians. Once a set of evolution bases is fixed, the superposition-free operations (SFOs) and superposition-free super-operations (SFSOs) are characterized. Then we introduce measures to quantify superposition present in an evolution. The superpositions can be used as a resource to perform certain tasks, which are otherwise impossible by means of SFOs and SFSOs. To demonstrate that we identify maximally superposed evolutions possessing maximum resource, and show how these operations can be used to deterministically implement arbitrary quantum operations and super-operations. In multi-party evolutions, non-local superposition  across the parties and their possible quantification are presented. We briefly discuss the role of local and non-local superpositions in context of a-causality, temporal Bell correlation, and indefinite temporal and causal orders.      

There are certain differences between superposition between evolutions and superposition between states. An arbitrary superposition in operation bases may not lead to a valid resultant evolution, unlike in quantum states.
For example $\frac{1}{2}[\mathbb{I} + \sigma_x + \sigma_y + \sigma_z ]$ is not a valid (complete) operation, although each basis (Pauli matrix) leads to unitary evolution! Therefore, there have to be restrictions, either on the coefficients or on the reference bases, or even on both. 
We also remark that an arbitrary superposition of unrestricted unitaries could be made to give rise to another unitary if we let the system to weakly interact with an ancilla and then go through a post-selection \cite{Aharonov90}. Apart from this seemingly different feature, the superpositions in evolutions share several interesting common properties with superposition in states -- for example, it can be collapsed or projected. In Appendices \ref{App:CreateSupEv} and \ref{App:EvCollapse}, we discuss on creation of superposition between evolutions and collapse of evolution to a chosen one, respectively.

Throughout this work, we stick to finite  dimensional Hilbert spaces. In general, quantum operations (or super-operations) could transform quantum states with different Hilbert space dimensions. However, we assume either equal input and output Hilbert spaces, or consider the one with larger dimension. 

\

\section{Quantifying superposition}
      
We first start by defining a set of mutually orthogonal bases that would be considered as the reference bases for evolutions.
Any evolution or operation composed of a pure and probabilistic mixture of this set of reference bases is defined to be \emph{superposition-free} quantum operations (SFOs). In the same spirit, a quantum super-operation (see Appendix \ref{App:QuntEvol}) that transforms an arbitrary superposition-free operation to another superposition-free operation is classified as the \emph{superposition-free super-operations} (SFSOs). An operation that is not an SFO involves superposition between the reference bases, by definition. 

\

\begin{figure}[t]
\includegraphics[width=0.8 \columnwidth]{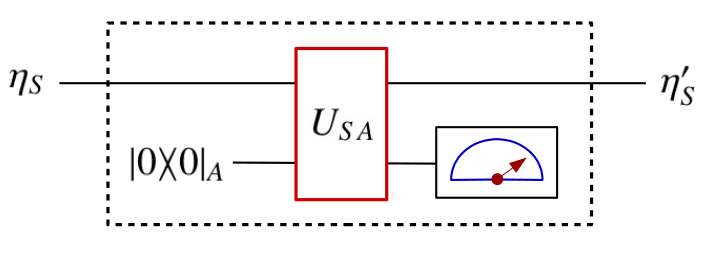}
\caption{A quantum operation on a quantum state $\eta_S$, say $\Phi$, is expressed using \emph{operator-sum-representation} \cite{Nielsen00}, as $\eta_S \longrightarrow \eta_S^\prime = \Phi(\eta_S)=\sum_m E_m \eta_S E_m^\dag$, where the $E_m$s are known as {\it operation elements} (also {\it Kraus operators}) and satisfy the condition $\sum_m E_m^\dag E_m \leqslant \mathbb{I}$. It can be implemented using quantum circuit, as depicted above, by (i) attaching system $\eta_S$ with an ancilla in a state $\proj{0}_A$, (ii) applying a global unitary $U_{SA}$ on the joint system, then (iii) projecting or tracing out ancilla, such that $\Phi(\eta_S)= \tr_A \left[ U_{SA} ( \eta_S \otimes \proj{0}_A ) U_{SA}^\dag \right]$. Note that the operation elements are $E_m=\bra{m}_A U_{SA}\ket{0}_A$. Deterministic implementation of operation, $\sum_m E_m^\dag E_m = \mathbb{I}$, corresponds to completely tracing out the ancilla $A$. On the other hand, selective projection on $A$-part leads to probabilistic (or selective) implantation of the operation, for which  $\sum_m E_m^\dag E_m < \mathbb{I}$. The selective implementation of operation element $E_m$ results in a quantum state $E_m \eta_S E_m^\dag/p_m$ with the probability $p_m= \tr (E_m \eta_S E_m^\dag)$.
\label{fig:QuantOp}}
\end{figure}

\noindent {\it Reference bases of quantum operations -- }  For a $d$ dimensional Hilbert space, there exists a set of $d^2$ mutually orthonormal Hilbert-Schmidt operators $\{ F_i \}$, which can be exploited as bases for arbitrary quantum operations.   The orthonormality means that  Hilbert-Schmidt norm satisfy $\tr \left(F_i^\dag F_j  \right)=d \delta_{ij} $. Note that there exists arbitrarily large number of basis sets, which are inter-related through unitary. For example, a set of $d^2$ mutually orthonormal non-unitary bases is given by $\{\sqrt{d} \ketbra{k}{l} \}$, where $\braket{k|l}=\delta_{kl}$ and $k,l=0, \ldots, d-1$. One may also we consider $d^2$ mutually orthonormal unitaries $\{U_i \}$
as the reference bases. 
In a qubit Hilbert space $(d=2)$, one can choose  $\mathbb{I} \cup \{\sigma_x, \sigma_y, \sigma_x \}$ as the reference bases, where $\sigma_i$s are the Pauli matrices.
For an arbitrary finite 
dimensional Hilbert space, a general orthonormal unitary basis set can be formed \cite{Schwinger60, Werner01, Oppenheim04} (see Appendix \ref{App:UniBasis}). 

The operations, corresponding to the reference bases, have an alternative representation in terms of {\it Choi-Jamiolkowski isomorphism} \cite{Jamiolkowski72, Choi75}, which relates input-output Hilbert spaces. Say the input and output Hilbert spaces are $\mathcal{H}_I$ and $\mathcal{H}_O$ respectively. Then a quantum operation corresponds to a set of reference bases $F_i$ on a state in  $\mathcal{H}_I$ produces an output state in $\mathcal{H}_O$ and the operation is fully characterized by a matrix in $\mathcal{H}_I \otimes \mathcal{H}_O$,
\begin{align}
 C_{F_i}= \proj{\psi_i},
\end{align}
where $\ket{\psi_i}:= [\mathbb{I} \otimes F_i] \ket{\psi}$, and $\ket{\psi}=\frac{1}{\sqrt{d}}\sum_k^d \ket{kk}$ is a maximally entangled state shared between input and output Hilbert spaces, each with  dimension $d$. Here $\ket{kk}= \ket{k}_I \otimes \ket{k}_O$ and $\{  \ket{k}_{I/O}\} \in \mathcal{H}_{I/O}$ are some mutually orthonormal state vectors, and $\mathbb{I}$ is the identity operation. The $C_{F_i}$ is also known as the {\it Choi matrix} corresponds to the operation  basis $F_i$. The condition $\tr \left(F_i^\dag F_j  \right)=d \delta_{ij}$ implies $\langle \psi_i | \psi_j \rangle= \delta_{ij}$. We denote $\ket{\psi_i}$ as the Choi state corresponds to the basis operation  $F_i$. Clearly, choosing $\{ F_i\}$ as the reference basis for quantum operation becomes equivalent to choosing $\{\ket{\psi_i} \}$ as reference basis states at the level of Choi matrix. 

\

\noindent {\it Superposition-free quantum operation -- } A general quantum operation can be expressed in terms of \emph{operator-sum-representation} \cite{Nielsen00} (see Fig.~\ref{fig:QuantOp}). 
Now, for a given $d^2$ reference bases $\{F_i\}$ acting on a $d$-dimensional Hilbert space, a quantum operation $\Lambda^F$ is defined to be  superposition-free if there exists an operation-sum-representation, with the operation elements $\{L_m\}$, such that
\begin{align}
 L_m \propto F_i, \  \  \ \forall m,
\end{align}
where $i=0 \ldots d^2-1$ and $\sum_m L_m^\dag L_m \leqslant \mathbb{I}$. Therefore the action of $\Lambda^F$ on an arbitrary quantum state $\eta$ becomes
\begin{align}
\label{eq:sfo-ru}
 \Lambda^F(\eta)=\sum_m L_m \eta L_m^\dag = \sum_{i=0}^{d^2-1} p_i \ F_i \eta F_i^\dag,
\end{align}
where $1 \geqslant p_i \geqslant 0$ and $\sum_i p_i \leqslant 1$. Clearly, a superposition-free operation is a probabilistic mixture of the reference operations where the operation elements are chosen from the set $\{F_i\}$. The fulfillment of the condition $\sum_m L_m^\dag L_m = \mathbb{I}$ (and $\sum_i p_i = 1$) implies deterministic implementation of the operation $\Lambda^F$. Any probabilistic or selective implementation will lead to  $\sum_m L_m^\dag L_m < \mathbb{I}$ (and $\sum_i p_i < 1$). The superposition-free quantum evolutions form a convex set, i.e. $\Lambda^F \in \mathcal{F}$, since any probabilistic mixture of superposition-free quantum evolutions is also a superposition-free quantum evolution. 

These operations can also be understood in terms of Choi matrices. Any evolution $\Lambda^F \in \mathcal{F}$ is superposition free {\it iff} it's corresponding Choi matrix is diagonal in reference Choi-states $\{\ket{\psi_i}\}$, i.e.,
\begin{align}
\label{eq:Choi-states}
 C_{\Lambda^F}=[\mathbb{I}\otimes \Lambda^F] \left(\proj{\psi}\right)= \sum_{i=0}^{d^2-1} p_i \proj{\psi_i},
\end{align}
with $1 \geqslant p_i \geqslant 0$ and $\sum_i p_i \leqslant 1$. The quantum operation $\Lambda^F$ is completely positive and trace preserving if and only if $C_{\Lambda^F} \geqslant 0$ and $\Tr_O [C_\Phi]=\frac{\mathbb{I}}{d}$ respectively. If $\Lambda^F$ represents a probabilistic or selective implementation of an operation, then $C_{\Lambda^F} \geqslant 0$ and $\Tr_O [C_\Phi]<\frac{\mathbb{I}}{d}$. 
The Choi matrices $C_{\Lambda^F}$, corresponds to $\Lambda^F \in \mathcal{F}$, also form a convex set and we denote it as $\mathcal{F}_{C}$, i.e., $C_{\Lambda^F} \in \mathcal{F}_{C}$. 

\

\begin{figure}
\includegraphics[width=0.9 \columnwidth]{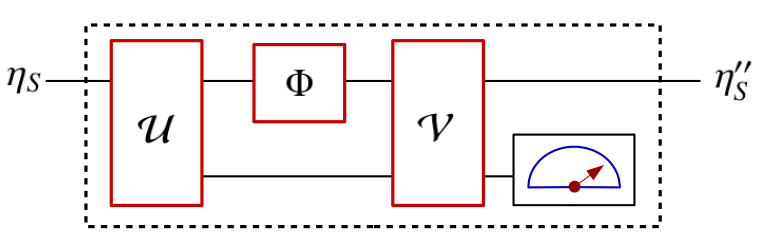}
\caption{A quantum super-operation \cite{Chiribella08} can be implemented using quantum isometry operations \cite{Iten16} in a quantum circuit. A  super-operation $\tilde{\Omega}$, that transforms an initial $\Phi$ to a final $\Lambda$ quantum operation, is designed as $\Lambda (\eta_S)=\tilde{\Omega}(\Phi)(\eta_S) = \tr_B \left[ V (\Phi \otimes \mathbb{I}_B) (U \eta_S U^\dag ) V^\dag \right]$. It is realized by two operations $\mathcal{U}=U \cdot U^\dag$ and $\mathcal{V}=V \cdot V^\dag$, where $U$ and $V$ are the isometries applied before and after the operation $\Phi$. At the end of the circuit, the ancillary system $B$ is either measured (using projective measurements), where each outcome results in a probabilistic (or selective) super-operation, or simply discarded (traced out), thereby realizing a deterministic super-operation. 
\label{fig:QuantSupOp}}
\end{figure}

\noindent {\it Superposition-free quantum super-operations -- } Similar to quantum states, a quantum operation can be transformed to result in a new quantum operation. These transformations are known as quantum super-operations \cite{Chiribella08} (see Fig.~\ref{fig:QuantSupOp} and  Appendix \ref{App:QuntEvol}). 
A super-operation is defined to be superposition-free, in reference to the set of bases $\{F_i\}$, if it transforms an arbitrary superposition-free operation to another superposition-free operation. In other words, a super-operation $\tilde{\Omega}^F$ is superposition-free, if  
\begin{align}
\label{eq:sfso}
 \Lambda^F= \tilde{\Omega}^F \left( \Phi^F \right) \in \mathcal{F}, \ \ \  \forall \ \Phi^F \in \mathcal{F}.
\end{align}
Operationally these super-operations transform a probabilistic mixture of basis operations to another probabilistic mixture of basis operations, where the bases are only chosen from reference set $\{F_i \}$, as in Eq.~\eqref{eq:sfo-ru}. These super-operations can be restricted further in the sense that, if they are applied selectively (i.e. exploiting probabilistic super-operations), they should still transform an arbitrary superposition-free operation to another superposition-free one. Though it is mathematically tedious to represent, we can easily express these super-operations in terms of Choi matrices (see  Appendix \ref{App:QuntEvol}). Then the $\tilde{\Omega}^F$ can equivalently be expressed with $\Omega^F$, which has an operator-sum-representation. It relates input $C_{\Phi^F}$ and output $C_{\Lambda^F}$ Choi matrices (see Eq. \ref{eq:sokr}), correspond to input $\Phi^F$ and output $\Lambda^F$ operations, respectively.  The  Eq.~\eqref{eq:sfso} can be recast as 
\begin{align}
\label{eq:sfsokr}
C_{\Lambda^F}= \Omega^F \left(C_{\Phi^F} \right)= \sum_n S_n^F C_{\Phi^F} S_n^{F \dag}, 
\end{align}
where $S_n^F$ are the super-operation elements acting on the input Choi matrix $C_{\Phi^F}$ to give rise to the output Choi matrix $C_{\Lambda^F} $. For a deterministic super-operation, we have $ \sum_n S_n^{F \dag} S_n^F = \mathbb{I}$. Furthermore, every deterministic super-operation guarantees that $\tr_O [C_{\Lambda^F}] =\frac{\mathbb{I}}{d}$, for all $C_{\Phi^F}$ with $\tr_O [C_{\Phi^F}] =\frac{\mathbb{I}}{d}$.  In this representation, we can also express probabilistic super-operation that includes selective or incomplete implementation of the super-operations, where $ \sum_n S_n^{F \dag} S_n^F < \mathbb{I}$. This also implies that $\tr_O [C_{\Lambda^F}]<\frac{\mathbb{I}}{d}$, for some $C_{\Phi^F}$ with $\tr_O [C_{\Phi^F}] =\frac{\mathbb{I}}{d}$. A selective implementation of a super-operation element $S_n^F$ will result in a selective super-operation $\Omega_n^F$, corresponds to a Choi matrix $C_{\Lambda^F_n}= S_n^F C_{\Phi^F} S_n^{F \dag}/q_n$, with a probability $q_n=\tr\left( S_n^F C_{\Phi^F} S_n^{F \dag}\right)/\tr (C_{\Phi^F})$. The resultant selective super-operation  $\tilde{\Omega}_n^F$ can then be derived from the Choi matrix $ C_{\Lambda^F_n}$.  Now to assure $\Omega^F$ (also $\tilde{\Omega}^F$) to be an SFSO, we impose the stricter condition, that is, for all $n$, 
\begin{align}
\label{eq:sfsokr-cond}
 S_n^F C_{\Phi^F} S_n^{F \dag} \in \mathcal{F}_{C}, \ \ \ \forall \ C_\Phi^F \in \mathcal{F}_{C}.
\end{align}
This means every super-operation element is constrained to give rise to a transformation corresponds to a probabilistic mixture of reference bases operations. Evidently, this condition for SFSOs is much stronger than that of Eq.~\eqref{eq:sfso}, as the former guarantees the latter but the converse is not necessarily true. In this work, we shall adhere to this stricter condition, given in Eq.~\eqref{eq:sfsokr-cond}. The SFSOs, $\tilde{\Omega}^F$ (and $\Omega^F$), form a convex set and we denote it as $\mathcal{O}$ (and $\mathcal{O}_{C}$).

In Choi matrix based representation, the structure of SFOs and SFSOs are similar to incoherent states and incoherent operations (superposition-free states and superposition-free operations) in the resource theory of quantum coherence \cite{Baumgratz14} (superposition \cite{Theurer17}). Except the fact that there are certain restrictions on the superposition coefficients in case of operations. However, the characterization and quantification of superposition as a resource, and their role in general transformations of operation and super-operations can be followed from \cite{Baumgratz14, Chitambar16, Theurer17, Dana17}.  

\

\noindent {\it Measure of superposition between quantum evolutions -- } We now introduce a class of functionals $M$ that maps quantum operation to a non-negative real number and satisfy some desirable properties to be a reliable quantifier of superposition. We demand the following criteria.

\noindent
(c1) $M(\Lambda^F)=0, \ \forall \Lambda^F \in \mathcal{F}$, i.e., the measure should give zero for all superposition-free quantum operations. One could consider a stronger condition that is (c1a) $M(\Lambda^F)=0$ iff $\Lambda^F \in \mathcal{F}$, and guarantees that $M(\Phi) > 0$ if $\Phi \notin \mathcal{F}$. Clearly, condition (c1a) implies (c1). 

\noindent
(c2a) $M$ monotonically decreases under SFSOs, i.e., $M(\Phi) \geqslant M \left( \tilde{\Omega}^F(\Phi) \right)$ for all $\tilde{\Omega}^F \in \mathcal{O}$. However, this cannot capture the monotonic property for selective or probabilistic SFSOs. To incorporate such cases, a stronger condition is necessary; (c2b) $M$ monotonically decreases under selective or probabilistic SFSOs on average, i.e., $M(\Phi) \geqslant \sum_n q_n M(\Phi_n)$. Here $\Phi_n$ are the operations as the results of selective SFSOs, which are implemented with the probabilities $q_n$.

\noindent
(c3) $M$ is non-increasing under mixing of quantum operations. This is also known as convexity condition, i.e., $\sum_n r_n M(\Phi_n) \geqslant  M\left(\sum_n r_n \Phi_n \right)$, for any set of quantum operations $\{\Phi_n \}$ with $ 1 \geqslant r_n \geqslant 0$ and $\sum_n r_n=1$. 

A measure that fulfills conditions (c2b) and (c3), satisfies (c2a) \cite{Baumgratz14}. Now we move on to propose some functionals that are potential measures for superpositions. Quality of any such measure, in fact, boils down to its capability to discriminate arbitrary quantum operation from the superposition-free ones.
In general, discriminating quantum operations are complex compare to quantum states, and one of the most useful approach is to exploit distance based measures. Any distance $D(\Phi, \Lambda)$ that could discriminate between quantum operations $\Phi$ and $\Lambda$, in a meaningful sense, should satisfy certain physically motivated criteria \cite{Gilchrist05}. (d1) {\it Metric:} This requires three important properties, (d1a)  $D(\Phi, \Lambda) \geqslant 0$ and $D(\Phi, \Lambda)=0$ iff $\Phi=\Lambda$, (d1b) symmetric, $D(\Phi, \Lambda)=D(\Lambda, \Phi)$ and (d1c) satisfies the triangular inequality, $D(\Phi, \Lambda) \leqslant D(\Phi, \Sigma) + D(\Sigma, \Lambda)$. (d2) {\it Stability:} $D(\Phi \otimes \mathbb{I}, \Lambda \otimes \mathbb{I})= D(\Phi, \Lambda)$, where identity operation $\mathbb{I}$ is applied on an additional quantum system. This physically means that an unrelated additional quantum system does not affect the value of $D$. (d3) {\it Unitary invariance:} $D(U \circ \Phi \circ V, U \circ \Lambda \circ V)= D(\Phi, \Lambda)$, where $U$ and $V$ are unitary operations. (d4) {\it Chaining:} $D(\Phi_2 \circ \Phi_1, \Lambda_2 \circ \Lambda_1) \leqslant D(\Phi_1, \Lambda_1) + D(\Phi_2, \Lambda_2)$. Physically it means that, for a process composed of many smaller processes, the combined distance is smaller than the sum of distances of the small processes. An important property like {\it contractivity}, i.e., $D(\Sigma \circ \Phi, \Sigma \circ \Lambda) \leqslant D(\Phi, \Lambda)$, where $\Sigma$ is any quantum operation, can be followed from (d1) and (d4). 

Easy to calculate distances, that fulfill all these criteria, are often based on Choi matrices. For two quantum operations, $\Phi$ and $\Lambda$, with  corresponding Choi matrices $C_{\Phi}$ and $C_{\Lambda}$ respectively, they are defined as $D_{C}(\Phi, \Lambda)= D(C_\Phi, C_\Lambda) $.
A class of distance based measures of superposition are then given by  
\begin{align}
 M \left(\Phi \right):= \min_{\Lambda^F \in \mathcal{F}} D_{C} \left(\Phi, \Lambda^F \right),
\end{align}
which represents the minimal distance of $\Phi$ to the set of superposition-free operations $\mathcal{F}$. 
We propose two such measures, below.

\noindent (1) The relative entropy of superposition
\begin{align}
\label{eq:Mr}
 M_r(\Phi)=\min_{C_{\Lambda^F} \in \mathcal{F}_{C}} S\left(C_\Phi \parallel C_{\Lambda^F} \right),
\end{align}
where the relative entropy is defined as $S(\rho \parallel \sigma)=\tr \left(\rho \ln \rho - \rho \ln \sigma   \right)$. Note that relative entropy respects all the qualities to be a good distance measure except symmetric condition (d1b). However, it has operational meaning from the perspective of information theory. It can be shown that the superposition-free operation for which the minimum in Eq.~\eqref{eq:Mr} is achieved, is $C_{\Phi^F}=\sum_i \ \braket{\psi_i|C_{\Phi}|\psi_i} \ \proj{\psi_i}$. Recall, $\{ \ket{\psi_i} \}$ are the Choi-states correspond to the bases $\{ F_i \}$.

\noindent (2) The $l_1$-measure of superposition
\begin{align}
 M_{l_1} (\Phi)=\min_{C_{\Lambda^F}} | C_{\Phi} - C_{\Lambda^F}|_1,
\end{align}
where $|\rho|_1=\sum_{ij} |\braket{\psi_i|\rho|\psi_j}|$, and $\{ \ket{\psi_i} \}$ are the Choi-states correspond to the reference bases $\{ F_i \}$. Again, the minimum is achieved for  $C_{\Phi^F}=\sum_i \ \braket{\psi_i|C_{\Phi}|\psi_i} \ \proj{\psi_i}$ and then $M_{l_1} (\Phi)= \sum_{i\neq j} |(C_{\Phi})_{ij}|$, 
where $C_{\Phi}=\sum_{ij} (C_{\Phi})_{ij} |\psi_i \rangle \langle \psi_j |$. 
One could also introduce other measures, similar to the ones used for the quantification of coherence and superposition in quantum states \cite{Streltsov17}, based on rank \cite{Theurer17}, trace-distance \cite{Rana16}, robustness \cite{Napoli16}, max-entropy \cite{Bu17} etc, which we shall not consider here. 

\

\noindent {\it Maximally-superposed operations -- } We have indicated that the presence of superposition in quantum evolutions is a resource when we are restricted to set of superposition-free operations and super-operations. In order to understand that, we first identify operations with the maximal resource,
acting on $d$-dimensional Hilbert space $\mathcal{H}_d$. For a $d^2$ reference bases $\{F_i \} $, the maximal resource operations are the (pure) unitary operations, $\sigma \rightarrow \eta = U_{\max} \sigma U_{\max}^\dag$, and these are given by
\begin{align} \label{eq:max-sup-op}
 U_{\max}=\frac{1}{d} \sum_{i=0}^{d^2-1} f_i F_i, 
\end{align}
where the complex coefficients $|f_i|=1, \ \forall i$.
In terms of Choi states, it becomes $\ket{\psi_{\max}}=\frac{1}{d} \sum_i f_i \ket{\psi_i}$. These are also called the maximally superposed operations. The resource measures achieve maximum value for these operations.
For example, in qubit ($d=2$) Hilbert space and given the reference bases $\mathbb{I} \cup \{\sigma_x, \ \sigma_y, \ \sigma_z \}$, a maximally superposed quantum operation is $U_{\max}=\frac{e^{i \phi}}{2} \left[ \mathbb{I} + i (\sigma_x + \sigma_y + \sigma_z)  \right]$. 

These operations have the maximum resource in the sense that, by means of superposition-free super-operations, any quantum operation $\Phi$ acting on $\mathcal{H}_d$ can be deterministically generated from them, i.e.,
\begin{align}
 \Phi = \tilde{\Omega}^F (U_{\max}),
\end{align}
where the arbitrary operation is $\Phi(\rho)=\sum_m E_m \rho E_m^\dag$ with the operation elements $E_m=\sum_i c_{mi} F_i$. Also, these maximally superposed operations, when consumed, allow one to deterministically implement arbitrary quantum super-operations, i.e.
\begin{align}
\tilde{\Omega}^F \left( \Phi \otimes U_{\max}  \right) \ \longrightarrow \ \tilde{\Omega} (\Phi),
\end{align}
where $\tilde{\Omega}^F$ is a superposition-free super-operation and $\tilde{\Omega}$ is an arbitrary super-operation. 
Explicit constructions of these processes are outlined in Appendices \ref{App:ImpOp} and \ref{App:ImpSupOp}. 

\section{Discussion}
Superposition plays pivotal roles in quantum mechanics. In the level of states, it is necessary for a system to exhibit ``quantum'' features. Quantum mechanics also allows superpositions between quantum evolutions. In this work, we have studied such superposition in the framework of resource theory. This allows us not only to quantify superposition but also enables us to identify maximally superposed evolutions and show how these evolutions can be exploited as resource performing certain quantum tasks which are otherwise impossible. 

There are several similarities and dissimilarities between the superpositions of states and evolutions. For example, both exhibit ``wave''-like behavior, in the sense that they can be selectively collapsed to a particular state or evolution. On the other hand, while an arbitrary superposition among quantum states could represent another state, this is not true for evolutions. Therefore, superpositions between evolutions are more restrictive, compared to states. Another important difference at the level of bases is that, while non-orthogonal quantum states are not perfectly distinguishable, non-orthogonal unitaries can be made perfectly distinguishable if one has access to finite copies \cite{Acin01}. Orthonormality, in the sense of Hilbert-Schmidt norm, only guarantees that each reference evolution basis can be perfectly distinguished from others, at one-copy level. Information, as like in  states, can be stored in operations too. However, information compression in operations are very different compared to the states and final operation with compressed information cannot be expressed in a separable form \cite{Oppenheim04}. Therefore performing important information theoretic tasks such as distillation, dilution, formation of superposition of operations etc, will be different from its quantum state analogs. These are interesting problems and are left open for future work. 

Evolutions that are applied to multi-party systems may have superpositions coming out of two different origins. These are local and non-local superpositions. The local ones are those that exclusively depend on the local bases and can be made vanishing by suitably choosing the local evolution bases. On the other hand, the non-local superposition cannot be made vanishing for any choice of local evolution bases. Superposition, in particular its non-local part, plays the most important role to exhibit non-trivial correlations in space-like separated quantum states. It is thus expected that superposition between evolutions would be instrumental to manifest non-trivial quantum phenomena between time-like separated quantum events, e.g. quantum a-causality, indefinite temporal and causal orders, and temporal Bell correlations. With this goal, we have studied the non-local superposition in an arbitrary bipartite quantum evolution in Appendix~\ref{App:corr}. The non-local superposition of an operation is closely connected to the quantum correlation generated when the operations is applied to a bipartite quantum state. Note, all the quantum properties of evolutions are not yet understood in fullest scales. Although, there are few proposals for experimentally realizable quantum operations that are able to exhibit these quantum behaviors. We have considered these operations in Appendix~\ref{App:acausal} and  explored the roles of non-local superpositions. We have shown that indeed superposition between evolutions is a necessary to have indefinite temporal order, and also the non-local superpositions are necessary ingredient for an operation to exhibit temporal Bell correlation and indefinite causal order. Interestingly, an operation does not require superposition to exhibit a-causal behavior. Note, these conclusions, on the roles of superposition between evolution in manifestations of non-trivial quantumness, are far from complete. However, the systematic understanding of superpositions presented here, in a framework of resource theory, may enable us to study these quantum phenomena from an alternative approach.      

\section{Acknowledgement}
The author acknowledges financial supports from Max-Planck Institute fellowship, Spanish Ministry MINECO (National Plan 15 Grant: FISICATEAMO No. FIS2016-79508-P, SEVERO OCHOA No. SEV-2015-0522, FPI), European Social Fund, Fundaci\'o Cellex, Generalitat de Catalunya (AGAUR Grant No. 2017 SGR 1341 and CERCA/Program), ERC AdG OSYRIS, EU FETPRO QUIC, and the National Science Centre, Poland-Symfonia Grant No. 2016/20/W/ST4/00314.



\appendix


\section{Quantum evolutions \label{App:QuntEvol}}

\noindent {\it Quantum operations -- }
Quantum operations describe all possible physical processes a quantum system may go through \cite{Nielsen00}, including unitary evolutions, quantum measurements, decoherences, and noise. A general quantum operation $\Phi$ may be expressed in {\it operator-sum representation} (also known as {\it Kraus representation}) relating input $\sigma$ and output $\eta$ states,
\begin{align}
 \eta= \Phi (\sigma)= \sum_m E_m \sigma E_m^\dag,
\end{align}
where the $E_m$s are known as {\it operation elements} (also known as {\it Kraus operators}) and satisfy the condition $\sum_m E_m^\dag E_m \leqslant \mathbb{I}$. The effect of the process is completely described by the operation elements $\{ E_m\}$. A trace-preserving operation that preserves trace of input and output states, fulfills $\sum_m E_m^\dag E_m = \mathbb{I}$. This physically means that the process described by $\Phi$ does not include post-selection and such an operation is traditionally called as {\it completely positive trace preserving} (CPTP) operation \cite{Nielsen00}. This also represents the situation where one does not have access to the individual outcomes of operation elements $E_m$. Incorporation of post-selection leads to trace-non-increasing operations and then $\sum_m E_m^\dag E_m < \mathbb{I}$. Selective implementation of the operation element $E_m$ results in a quantum state $\eta_m=E_m \sigma E_m^\dag / p_m$ with the probability $p_m=\tr(E_m \sigma E_m^\dag)$. In fact, we could write $ \eta = \sum_m \Phi_m (\sigma)= \sum_m p_m \eta_m$.

Note, the operation elements do not necessarily have to be square matrices. In the case where the dimensions of input $\mathcal{H}_d$ and output $\mathcal{H}_{d^\prime}$ Hilbert spaces are different, the $E_m$s are represented by  $d^\prime \times d$ matrices. However, one could always choose square matrices of  $d \times d$ as operation elements, where $d \geqslant d^\prime$.

A drawback of operator-sum representation of a quantum operation is that it is not unique, in the sense that there are many sets of operation elements to give rise to the same operation \cite{Nielsen00}. Therefore comparison between quantum operations become difficult. To alleviate such a problem, one may fix a set of bases $\{F_i \}$ in the space of operators that are mutually orthonormal under Hilbert-Schmidt inner product, i.e., $\tr\left( F_i^\dag F_j \right)=d \delta_{ij}$. These orthonormal bases can be used to expand the operation elements as $E_m=\sum_i a_{mi} F_i$, where $A=\{ a_{mi}\}$ is a unitary matrix \cite{Nielsen00}. Now the operation is recast as
\begin{align}
\label{eq:KrausRef}
 \eta= \Phi (\sigma)= \sum_{ij} \left( \xi_{\Phi} \right)_{ij} F_i \sigma F_j^\dag,
\end{align}
where $\left( \xi_{\Phi} \right)_{ij}\equiv \sum_m a_{im}a_{mj}^\star$ are the elements of the {\it process matrix} $\xi_{\Phi}$. Clearly, once the set of orthonormal operator bases is fixed, the process matrix becomes unique to the process, i.e., only on $\Phi$ and not on the choice of $E_m$, and completely describes the quantum operations \cite{Gilchrist05}. For non-trace-preserving quantum operations ($ \sum_m E_m^\dag E_m < \mathbb{I}$) the process matrix becomes $\sum_{ij} \left( \xi_{\Phi} \right)_{ij} F_j^\dag F_i <\mathbb{I}$.

\

Another, closely related, but more abstract representation of a quantum operation can be given in terms of {\it Choi-Jamiolkowski isomorphism} \cite{Jamiolkowski72, Choi75}, which related input-output Hilbert spaces for an arbitrary quantum operation. Considering input and output Hilbert spaces $\mathcal{H}_I$ and $\mathcal{H}_O$, where a quantum operation $\Phi$ is applied on a state in  $\mathcal{H}_I$ to produce an output state in $\mathcal{H}_O$, the operation is fully characterized by a matrix in $\mathcal{H}_I \otimes \mathcal{H}_O$, given by
\begin{align}
C_{\Phi} \equiv [\mathbb{I}\otimes \Phi] ( \proj{\psi} ), 
\end{align}
where $\ket{\psi}=\frac{1}{\sqrt{d}}\sum_k^d \ket{kk}$ is a maximally entangled state in $d$-dimensional system Hilbert space with another copy of itself. Here $\ket{kk}= \ket{k}_I \otimes \ket{k}_O$ and $\{  \ket{k}_{I/O}\} \in \mathcal{H}_{I/O}$ are some orthonormal basis set, and $\mathbb{I}$ is the identity operation. The $C_{\Phi}$ is also known as the {\it Choi matrix} corresponds to the operation $\Phi$, and their one-to-one correspondence is given by 
\begin{align}
 \Phi(\rho)=\tr_I \left[(\rho^T \otimes \mathbb{I}) \ C_{\Phi}  \right].
\end{align}
The quantum operation $\Phi$ is completely positive and trace preserving if and only if 
\begin{align}\label{eq:cptp-cond}
 C_\Phi \geqslant 0 \ \ \mbox{and} \ \  \Tr_O [C_\Phi]=\frac{\mathbb{I}}{d},
\end{align}
respectively.
This isomorphism guarantees an equivalence between $\Phi$ and $C_{\Phi}$, and therefore it enables us to treat quantum operations with the tools that are ordinarily used to treat quantum states. 
For any completely positive but trace non-preserving operation one has $\sum_m E_m^\dag E_m < \mathbb{I}$, which includes probabilistic, partial or selective implementation of an operation, will have $\tr_O \left[C_{\Phi}  \right] < \frac{\mathbb{I}}{d}$. The process matrix $\xi_{\Phi}$ and the Choi matrix $C_{\Phi}$ are closely related to each other \cite{Gilchrist05}. However, it is mathematically convenient to use Choi matrix $C_{\Phi}$, which is considered in this work.

\

\noindent {\it Quantum super-operations -- }
While quantum operations describe quantum processes that may occur to a quantum system, there are also quantum {\it super-operation} that transforms one (input) quantum operation to another (output) quantum operation \cite{Chiribella08}. Quantum super-operation, in fact, constitutes the most general kind of transformations between elementary quantum objects. A general quantum super-operation $\tilde{\Omega}$ relates input $\Phi$ and output $\Lambda$ quantum operations as 
\begin{align}
\Lambda= \tilde{\Omega} \left( \Phi \right).
\end{align}
{\it Deterministic} super-operations transform a CPTP operations to another CPTP operations. Conversely, a {probabilistic} super-operation could transform a CPTP operation into a trace-non-increasing operation, which is due to a post-selection during the super-operation, like in operations. Physically an arbitrary quantum super-operation can be implemented with the help of quantum circuits \cite{Chiribella08}. However mathematical representation of the same is cumbersome, in caparison to quantum operations. 

One way to circumvent this difficulty is the use of Choi-Jamiolkowski isomorphism, by which a quantum operation can be equivalently expressed in terms of quantum state (Choi matrix). Then a super-operation, which induces transformations between quantum operations, can be understood as a ``quantum operation'' that relates input and  output Choi matrices, with an associated operator-sum-representation. For a super-operation, $\tilde{\Omega}$ relating input $\Phi$ and output $\Lambda$ operations, we may write
\begin{align}
\label{eq:sokr}
 C_{\Lambda}= \Omega \left( C_\Phi \right) =\sum_n S_n C_\Phi S_n^\dag,
\end{align}
where $C_{\Lambda}$ and $C_\Phi$ are the Choi matrices correspond to quantum operations $\Lambda$ and $\Phi$ respectively, and $S_n$s are the super-operation elements of $\Omega$ on the level of Choi matrix. The operation $\Omega$ has a one-to-one correspondence with the super-operation $\tilde{\Omega}$ \cite{Chiribella08}. A deterministic super-operation $\tilde{\Omega}$ means that $\Omega$ is not only a CPTP operation, i.e., $\sum_n S_n^\dag S_n = \mathbb{I}$, but also has to satisfy 
\begin{align}\label{eq:cond-supop}
 \tr_O \left( C_{\Lambda} \right) = \frac{\mathbb{I}}{d}, \ \ \ \forall \left\{ C_\Phi | \tr_O \left( C_{\Phi} \right) = \frac{\mathbb{I}}{d} \right\}.
\end{align}

In contrast, a probabilistic super-operation $\tilde{\Omega}$ means a trace-non-increasing operation $\Omega$ with $\sum_n S_n^\dag S_n < \mathbb{I}$, where some post-selections are done. Any selective implementation of super-operation element $S_n$ leads to a partial operation $\Omega_n (C_{\Phi})=S_n C_{\Phi} S_n^\dag$ that corresponds to an implementation of an operation $\tilde{\Omega}_n$ with respect to the Choi matrix $C_{\Lambda n}=S_n C_{\Phi} S_n^\dag / q_n$ with probability $q_n=\tr (S_n C_{\Phi} S_n^\dag)$. Note $C_{\Lambda}= \sum_n q_n  C_{\Lambda n}$. 
In general, the reduction of operation elements $S_n$ at the level of quantum super-operations can be done and there exist quantum circuits to implement both deterministic and probabilistic super-operations \cite{Chiribella08}.

\

\section{Reference bases \label{App:UniBasis}}
There are infinitely many ways one can choose the reference bases, which are mutually orthonormal in the sense of Hilbert-Schmidt norm, and all of these bases are inter-related through unitaries. For simplicity, we shall consider two choices of reference bases. One is a set of non-unitary bases and another one is a set of unitary bases. We also provide the unitary operation that inter-converts these two sets of bases.

\

\noindent {\it Non-unitary bases -- } There could be many choices of non-unitary orthonormal bases. For example, we may have $d^2$ mutually orthonormal bases, for $ i,j=0,1, \ldots, d-1$,
\begin{align}\label{eq:nu-basis}
\left\{R_{ij} \right\}=\left\{ \sqrt{d} \ \ketbra{i}{j}\right\},
\end{align}
where $\tr \left( R_{ij} R_{mn}^{\dag} \right)=d \delta_{im} \delta_{jn}$, $\forall i,j,m,n$. For qubit ($d=2$) Hilbert space, the set of reference bases becomes $\{\sqrt{2}\ketbra{0}{0}, \sqrt{2}\ketbra{0}{1}, \sqrt{2}\ketbra{1}{0}, \sqrt{2}\ketbra{1}{1}\}$. 

\

\noindent {\it Unitary bases -- } There are also many choices of sets of unitary bases. One of the constructions of orthonormal unitary bases for arbitrary finite dimensional Hilbert spaces goes back to the works by Schwinger \cite{Schwinger60}. For a $d$-dimensional case, a set of orthonormal bases are $\{ U_{mn} \}$, where subscripts take the values $m,n=0, \ldots , d-1$. Then 
\begin{align}
 U_{mn}=U_0 \ S_{mn},
\end{align}
where $U_0$ is a unitary and $\{ S_{mn} \}$ are $d^2$ orthonormal traceless unitary operators. The unitaries are constructed as \cite{Schwinger60, Oppenheim04}
\begin{align}\label{eq:Smn0}
 S_{mn} = Z^m X^n,
\end{align}
where the operators
\begin{align}\label{eq:Z}
 Z=\sum_{k=0}^{d-1} \xi^k \proj{k}
\end{align}
and 
\begin{align}\label{eq:X}
 X=\sum_{k=0}^{d-1}  \ket{(k+1) \ \mbox{mod} \ d} \bra{k},
\end{align}
satisfying $Z^d=X^d=\mathbb{I}$ and $ZX=\xi XZ$ with $\xi=\exp(2 i \pi/d)$. Note, the $Z$ and $X$ represents the generalized phase-flip and bit-flip operators respectively. For $d=2$, we immediately recover that $Z=\sigma_z$, $X=\sigma_x$ and consequently the set of orthonormal bases become $S_{00}=\mathbb{I}$, $S_{01}=\sigma_x$, $S_{10}=\sigma_z$ and $S_{11}=i \sigma_y$. 

These two sets of reference bases are inter-convertible through unitary operations.  
The set of non-unitary basis (in Eq.~\eqref{eq:nu-basis}) is related to the set of unitary bases (in Eq.~\eqref{eq:Smn0}) as
\begin{align}\label{eq:Nu-Smn}
R_{jk}= \sqrt{d} \ \ketbra{j}{k}= \sum_{l=0}^{d-1} a_{lj} \ S_{l [(k-j) \mod d]},
\end{align}
where $a_{lj}=\frac{1}{\sqrt{d}} e^{-\frac{2 \pi i}{d}lj}$, and $A=\{ a_{lj} \}$ is a unitary operator. Similarly, the set of unitary bases (in Eq.~\eqref{eq:Smn0}) is related to the set of non-unitary bases (in Eq.~\eqref{eq:nu-basis}) as
\begin{align}\label{eq:Smn}
 S_{mn}=\sum_{k=0}^{d-1} b_{nk} \ R_{k [(k+m)\mod d]},
\end{align}
for $m,n=0,1, \ldots, d-1$, where  $b_{kn}=\frac{1}{\sqrt{d}}e^{\frac{2 \pi i}{d}kn}$ and $B=\{b_{kn} \}$ is a unitary operator.

\section{Creating superposition between evolutions \label{App:CreateSupEv}}
Here we outline a protocol to create superposition between evolutions. Note, there may be many different ways to create such superpositions. One simple approach is the use of quantum switches. Consider a quantum system $S$ in a state $\ket{\psi_S}$, on which two unitaries $U_1$ and $U_2$ are applied. These two unitaries can be the results of time evolution driven by two Hamiltonians $H_1$ and $H_2$, i.e., $U_1=\exp[-iH_1 t]$ and $U_2=\exp[-iH_2t]$, where $t$ is time of evolution. There is also a two-level quantum switch $C$, which controls the unitary to be applied on the system. Say, if the switch is in $\ket{1}$ the $U_1$ is applied and similarly, $U_2$ is applied on the system if the switch is in state $\ket{2}$. Therefore the overall operation that is applied on the joint C-S system becomes
\begin{align}\label{eq:JCS}
 U_{CS}=\proj{1} \otimes U_1 + \proj{2} \otimes U_2.
\end{align}
Clearly, for a joint system $\proj{1} \otimes \proj{\psi_S}$, the unitary in Eq.~\eqref{eq:JCS} results in an implementation of $U_1$ on system $S$, $U_1 \proj{\psi_S} U_1^\dag$. Similarly, for $\proj{2} \otimes \proj{\psi_S}$, it results in $U_2 \proj{\psi_S} U_2^\dag$ on the system $S$. 

Now consider the quantum switch is in a state $\ket{\psi_C}$, which is in superposition between $\ket{1}$ and $\ket{2}$, $\ket{\psi_C}=c_1 \ket{1}+c_2\ket{2}$. Then the joint unitary on the joint C-S state gives rise to 
\begin{align}\label{eq:JCSApp}
 \ket{\psi_{CS}^\prime}&=U_{CS} \ket{\psi_c} \otimes \ket{\psi_S}, \nonumber \\
                       &=c_1 \ket{1} \otimes U_1 \ket{\psi_S} + c_2 \ket{2} \otimes U_2 \ket{\psi_S}.
\end{align}
Again if the switch is projected in $\proj{1}$ (and $\proj{2}$), the resultant unitary applied on the system would be $U_1$ (and $U_2$). Instead, if the switch is projected in a state which has superposition between $\ket{1}$ and $\ket{2}$, the resultant operation on the system would have a superposition between evolutions $U_1$ and $U_2$. For example, if the switch is projected in the state $\ket{+}=\frac{1}{\sqrt{2}}\left(\ket{1} + \ket{2} \right)$ (with the projector $\proj{+}$), then 
\begin{align}
 \ket{\psi_{CS}^{\prime \prime}}=\ket{+} \otimes  \left(\frac{c_1}{\sqrt{2}} U_1 + \frac{c_2}{\sqrt{2}} U_2 \right) \ket{\psi_S}.
\end{align}
So, up to a normalization constant, the projection leads to a superposition between evolutions driven by $U_1$ and $U_2$. 

Note, the protocol to create superposition can be extended to arbitrary evolutions and there the unitaries are replaced by general CPTP maps, $\Lambda_1$ and $\Lambda_2$.

\section{Collapsing evolution \label{App:EvCollapse}}
Here we shall consider quantum super-operations that collapse a superposed quantum evolution to one, selected evolution. This can be understood in analogy with collapse of a quantum state to a selected one using projective measurement.

We have mentioned earlier that there are certain differences between superposition between evolutions and superposition between states. An arbitrary superposition in operation bases may not lead to a valid  (i.e. physical) evolution, unlike in quantum states.
For example $\frac{1}{2}[\mathbb{I} + \sigma_x + \sigma_y + \sigma_z ]$ is not a valid (complete) operation, although each basis (Pauli matrix) leads to unitary evolution! Therefore, there has to have restrictions, either on the coefficients or on the reference bases, or even on both. 
Apart from this seemingly different feature, the superpositions in evolutions share a very interesting common property with superposition in states -- it can be collapsed or projected.

For simplicity, we give examples of unitary evolutions and choose sets of orthonormal unitary bases as reference bases. Extension to arbitrary evolution and arbitrary reference bases can be done easily.  
Consider an arbitrary (pure) unitary operation $V$ on an arbitrary quantum state $\eta$ in a $d$-dimensional Hilbert space, as $\eta  \rightarrow \ V \eta V^\dag$.
The unitary operator $V$ can be decomposed in terms of the orthonormal unitary bases $\{U_i\}$, as
\begin{align}
 V=\sum_{i=0}^{d^2-1} c_i U_i,
\end{align}
where amplitude $c_i=\frac{1}{d} \tr \left(U_i^\dag V \right)$,  $\sum_i|c_i|^2=1$ and $\sum_{i\neq j} c_i c_j^* U_iU_j^\dag=\sum_{i\neq j} c_i^* c_j U_i^\dag U_j=0$.
The parameters $c_i$s can also be derived from the Choi matrix corresponds to $V$, which satisfies $\tr_O\left(C_V\right)=\tr_I\left(C_V\right)=\frac{\mathbb{I}}{d}$. 
Now  a smart experimental protocol can be devised, where a measurement could lead to a selective collapse of the evolution 
to
\begin{align}
V \eta V^\dag \longrightarrow U_i \eta U_i^\dag.
\end{align}
As a result, only unitary $U_i$ is applied on the system, with the probability $|c_i|^2$ (see examples below). Note that this is equivalent to devise a super-projector (i.e. projector at the level of super-operation) and it is independent of the choice of the state $\eta$, on which the operation $V$ is applied. The same can also be performed for more general evolution, say  $\Phi(\eta)=\sum_m E_m \rho E_m^\dag$, where $E_m=\sum_i c_{mi} U_i$, in a similar fashion. 

\

\noindent {\it Collapse of evolution in qubit dimension ($d=2$) --} 
For simplicity, we consider evolution that maps qubit state to another qubit state and unitaries are expressed in a superposition of Pauli matrices. Consider a unitary 
\begin{align}
 V_S=c_0 \mathbb{I} + c_x \sigma_x + c_y \sigma_y + c_z \sigma_z, 
\end{align}
where $\mathbb{I}$ is the identity matrix and $\{\sigma_x, \sigma_y, \sigma_z \}$ are the Pauli matrices. We use them as our reference bases. However one may consider any set of orthonormal set of unitaries. To perform a collapse of the evolution, we follow the protocol given in \cite{Oppenheim04}.
Say the unitary $V_S$ to be applied on an arbitrary system state $\ket{\phi}_S$ and then steps to collapse the evolution are outlined as follows.

\

\noindent (1) The system is attached with two ancillary qubit systems $A$ and $B$ in states $\ket{+}_A$ and $\ket{+}_B$, where $\ket{+}_{A/B}=\frac{1}{\sqrt{2}}(\ket{0}_{A/B}+\ket{1}_{A/B})$ and $\{ \ket{0}, \ \ket{1}\}$ are the eigenstates of $\sigma_z$. Therefore the joint state becomes $\ket{+}_A \otimes \ket{+}_B \otimes \ket{\phi}_S$ in the Hilbert space $\mathcal{H}_{A}\otimes \mathcal{H}_{B}\otimes \mathcal{H}_S $. 

\

\noindent (2) We apply $U_B U_A$, where $U_{A}=\proj{0}_{A} \otimes \mathbb{I} + \proj{1}_{A} \otimes \sigma_z$ and $U_{B}=\proj{0}_{B} \otimes \mathbb{I} + \proj{1}_{B} \otimes \sigma_x$, and the $U_{A/B}$ are applied on the joint Hilbert spaces of the ancilla ($A/B$) and the system, $\mathcal{H}_{A/B} \otimes \mathcal{H}_S$. The resultant state becomes $U_B U_A \ket{+}_A \otimes \ket{+}_B \otimes \ket{\phi}_S$.

\

\noindent (3) Now we apply unitary $V_S$ on the system and then $U_A U_B$, such that
\begin{align}
  U_A  U_B V_S U_B U_A \left( \ket{++}_{AB} \otimes \ket{\phi}_S \right) =  \ket{\phi}_{ABS}.
\end{align}
The final state can also be written as 
\begin{align}
  \ket{\phi}_{ABS} =  \ \ & c_0 \ket{++}_{AB} \otimes \ket{\phi}_S \nonumber \\
                     & + c_1 \ket{+-}_{AB} \otimes \sigma_x \ket{\phi}_S \nonumber \\
                     & + c_2 \ket{--}_{AB} \otimes \sigma_y \ket{\phi}_S \nonumber \\
                     & + c_3 \ket{-+}_{AB} \otimes \sigma_z \ket{\phi}_S.
\end{align}

\

\noindent (4) Finally collapsing the evolution is equivalent to performing projective measurements on the ancillary systems $AB$ with the orthogonal projectors $\left\{ \proj{++}_{AB}, \ \proj{+-}_{AB}, \  \proj{--}_{AB}, \ \proj{-+}_{AB}  \right\}$ leading to reduced unitary evolutions $\{ \mathbb{I}, \sigma_x, \sigma_y, \sigma_z \}$ applied with the probabilities $\{ |c_0|^2, |c_1|^2, |c_2|^2, |c_3|^2 \}$. 

\

\noindent {\it Collapse of evolution in higher dimension ($d>2$) --}
Extension of collapse of an evolution beyond $d=2$ can be done as in the following. Consider a unitary 
\begin{align}
 V_S=\sum_{m,n=0}^{d-1} c_{mn} S_{mn},
\end{align}
that is acting on an arbitrary state $\ket{\phi}_S$ system in $d$-dimensional Hilbert space. Here we use the reference bases $ \{S_{mn}\}$, as expressed in Eq.~\eqref{eq:Smn0}. 
To perform measurement leading to collapse, we attach two ancillary systems ($A$ and $B$), each with $d$-dimensional Hilbert spaces and complete orthonormal state vectors $\ket{k}_A$ and $\ket{k}_B$, such that the joint state is $\ket{\psi_0}_A \otimes \ket{\psi_0}_B \otimes \ket{\phi}_S$, where $\ket{\psi_0}_{A/B}=\frac{1}{\sqrt{d}}\sum_{k=0}^{d-1} \ket{k}_{A/B}$. We introduce interactions between $A-S$ and $B-S$ with the operators  
\begin{align}
 U_A^Z=\sum_{k=0}^{d-1} \proj{k}_A\otimes Z^k,
\end{align}
and 
\begin{align}
 U_B^X=\sum_{k=0}^{d-1} \proj{k}_B\otimes X^k.
\end{align}
We apply $U_B^X U_A^Z$ before the unitary $V_S$ on the system, and then $U_A^{Z\dag} U_B^{X\dag}$, such that the resultant operation becomes
\begin{align}
 U_A^{Z\dag} U_B^{X\dag} \left( \sum_{m,n=0}^{d-1} c_{mn} S_{mn} \right) U_B^X U_A^Z. 
\end{align}
The action of this operation on the joint state can be written, with some manipulations, as
\begin{align}
 & \left[ U_A^{Z\dag} U_B^{X\dag}  \left( \sum_{m,n=0}^{d-1} c_{mn} S_{mn} \right) U_B^X U_A^Z \right] \ket{\psi_0}_A \otimes \ket{\psi_0}_B \otimes \ket{\phi}_S \nonumber \\
                    & = \sum_{m,n=0}^{d-1} c_{mn} \ \ket{\psi_m}_A \otimes \ket{\psi_n}_B \otimes S_{mn} \ \ket{\phi}_S,
\end{align}
where the ancilla states are mutually orthonormal to each other, i.e. $\braket{\psi_m | \psi_n}_{A/B}=\delta_{mn}$. Therefore a projective measurement on the ancillary systems with $\proj{\psi_m}_A \otimes \proj{\psi_n}_B$ will collapse the sum to a single term with the corresponding probability $|c_{mn}|^2$ with the unitary $S_{mn}$ applied on the system.  

\section{Maximally superposed evolution}
We have noted that an arbitrary superposition between the reference bases (or operation elements) does not give raise to a valid operation. There have to some restrictions on the coefficients or on the bases. With these constraints, we go on to show the operations with the maximum resource. The maximally superposed operations turn out to be unitary operations, irrespective to the choices of the reference bases. For a choice of a set of $d^2$ reference bases $\{F_i \}$, the maximally superposed operations can be expressed as
\begin{align}\label{eq:nu-mso}
 U_{\max}=\frac{1}{d} \sum_{i=0}^{d^2-1} f_i F_i,
\end{align}
where the complex coefficients $|f_{i}|=1$. Without loss of generality, we can choose a new set of reference bases, where $F_i^\prime = f_i F_i$, and then the corresponding maximally superposed operation becomes 
\begin{align}\label{eq:nu-mso1}
 U_{\max}=\frac{1}{d} \sum_{i=0}^{d^2-1} F_i^\prime.
\end{align}
In Appendices~\ref{App:ImpOp} and \ref{App:ImpSupOp}, we consider Eq.~\eqref{eq:nu-mso1} to study how these operations with the maximal resource can be exploited to implement arbitrary operations and super-operations.

\

In a $d$-dimensional Hilbert space and for the non-unitary reference bases $\{R_{kl} \}=\{ \sqrt{d} \ \ketbra{k}{l}\}$  (as shown in Eq.~\eqref{eq:nu-basis}), a maximally superposed operation is
\begin{align}\label{eq:nu-mso}
 U^{nu}_{\max}=\frac{1}{d} \sum_{k,l=0}^{d-1} r_{kl} R_{kl},
\end{align}
where the complex coefficients $r_{kl}=e^{-\frac{2 \pi i}{d} kl}$. Note that the $U^{nu}_{\max}$ is a unitary that corresponds to quantum (discrete) Fourier transformation operation. For simplicity, we can slightly modify the reference bases to $\{ R_{kl}^\prime = r_{kl} R_{kl}\}$ and in that case the maximally superposed operation becomes $U^{nu}_{\max}=\frac{1}{d} \sum_{k,l=0}^{d-1} R_{kl}^\prime$. For qubit Hilbert space ($d=2$), it reduces to the Hadamard gate,
\begin{align}\label{eq:Hadamard}
 U^{nu}_{\max}& =\frac{1}{2} \left( \sqrt{2} \proj{0} + \sqrt{2} \ketbra{0}{1} + \sqrt{2} \ketbra{1}{0} - \sqrt{2} \proj{1}  \right), \\ \nonumber 
 & = \frac{1}{\sqrt{2}}\begin{bmatrix}
  1 & 1 \\
  1 & -1
 \end{bmatrix}.
\end{align}

\

For unitary reference bases $\{S_{mn}\}$ (see Eq.~\eqref{eq:Smn0}) acting in a $d$-dimensional Hilbert space, a maximally superposed operation is 
\begin{align}\label{eq:u-mso}
 U^{u}_{\max}=\frac{1}{d} \sum_{k,l=0}^{d-1} s_{mn} \ S_{mn},
\end{align}
where complex coefficients $|s_{mn}|=1$. The complex coefficients can be derived from the $U^{nu}_{\max}$ in Eq.~\eqref{eq:nu-mso} and the unitary operation $B$ in Eq.~\eqref{eq:Smn}, as it inter-relates the reference bases, $B : \{ R_{ij} \} \longrightarrow \{ S_{mn} \}$.
Again for simplicity, we may slightly modify the reference bases to $ \{ S_{mn}^\prime = s_{mn} S_{mn} \}$ and then the maximally superposed operation becomes $U^{u}_{\max}=\frac{1}{d} \sum_{k,l=0}^{d-1}  S_{mn}^\prime $.
In case of qubit Hilbert space ($d=2$) and the reference bases $\mathbb{I} \cup \{i \sigma_x, i \sigma_y, i \sigma_z  \} $, a maximally superposed operation is given by
\begin{align}\label{eq:u-mso-2}
 U^{u}_{\max}=\frac{1}{2} \left( \mathbb{I} + i \sigma_x + i \sigma_y + i \sigma_z\right).
\end{align}

\section{Implementation of arbitrary operation using maximally-superposed operations \label{App:ImpOp}}
Here we show how an arbitrary quantum operation can be implemented by means of maximally-superposed operation, which is used as a resource, and superposition-free super-operations. The operations with maximum superposition is given in Eq. \eqref{eq:max-sup-op}. However, for the ease of derivations below, we slightly modify the set of $d^2$ reference bases as $\{F_i^\prime=f_i F_i \}$, for a $d$-dimensional Hilbert space. Then, as mentioned in Eq.~\eqref{eq:nu-mso1}, the maximally superposed operation becomes
\begin{align}\label{eq:Umax}
 U_{\max}=\frac{1}{d} \sum_i F_i^\prime.
\end{align}
In the Choi state representation these maximal resource operations become 
\begin{align}\label{eq:UmaxChoi}
 \ket{\psi_{\max}}= \frac{1}{d} \sum_i \ket{\phi_i},
\end{align}
where $\ket{\phi_i}=(\mathbb{I} \otimes F_i^\prime)\frac{1}{\sqrt{d}}\sum_k \ket{kk}$. We denote the Choi matrix corresponds to the maximally superposed operation as $C_{\max}=\proj{\psi_{\max}}$.

Let us first consider the implementation of an arbitrary unitary operation $V=\sum_{i=0}^{d^2-1} c_i F_i^\prime$, where $\sum_{i=0}^{d^2-1} |c_i|=1$ and $\sum_{i\neq j} c_i c_j^* F_i^\prime F_j^{\prime \dag} = \sum_{i\neq j} c_i^* c_j F_i^{\prime \dag} F_j^\prime=0$. The corresponding Choi state is $\ket{\psi_V}= \left(\mathbb{I} \otimes V \right) \frac{1}{\sqrt{d}}\sum_k \ket{kk}=\sum_{i=0}^{d^2-1} c_i \ket{\phi_i}$. At the level of Choi matrix, the implementation of the unitary reduces down to the generation of the Choi matrix
\begin{align}\label{eq:sfo-imp}
 \proj{\psi_V} & =\Omega^F\left(\proj{\psi_{\max}}  \right), \nonumber \\
               & =\sum_{n=0}^{d^2-1} S_n^F \left( \proj{\psi_{\max}} \right)S_n^{F\dag},
\end{align}
where $S_n^F$ elements of superposition-free super-operation. Such a super-operation can indeed be constructed following \cite{Baumgratz14}, where the super-operation elements are given by
\begin{align}\label{eq:ko-sfo}
 S_n^F=\sum_{i=0}^{d^2-1} c_i \ketbra{\phi_i}{\phi_{m_{i+n-1}}}, 
\end{align}
with $m_y=y- \left\lfloor \frac{y-1}{d^2} \right\rfloor d^2$. Note that these super-operation elements are ``strictly'' superposition-free and satisfy $\sum_n S_n^{F \dag}S_n^F=\mathbb{I}$. 

Now we turn to the implementation of an arbitrary quantum operation, by means of superposition-free super-operation and a maximally superposed operations. Consider an operation $\Phi(\rho)=\sum_m E_m \rho E_m^\dag$. The corresponding Choi matrix is $C_{\phi}=\sum_x p_x \proj{\psi_x}$. Each $\ket{\psi_x}$ can be expressed as $\ket{\psi_x}=\sum_{i=0}^{d^2-1} c_{xi} \ket{\phi_i}$. Now, similar to Eqs. \eqref{eq:sfo-imp} and \eqref{eq:ko-sfo}, we can devise a super-operation, such that
\begin{align}
 p_x \ \proj{\psi_x} & =\Omega^F_x\left(\proj{\psi_{\max}} \right) \nonumber \\
                     & = \sum_{n=0}^{d^2-1} S_{xn}^F \left( \proj{\psi_{\max}} \right)S_{xn}^{F\dag},
\end{align}
where 
\begin{align}
 S_{xn}^F=\sqrt{p_x} \ \sum_{i=0}^{d^2-1} c_{xi} \ketbra{\phi_i}{\phi_{m_{i+n-1}}}.
\end{align}
Now, the desired Choi matrix corresponds to $\Phi$ is achieved, as
\begin{align}
 C_{\phi}=\sum_{x,n} S_{xn}^F \left(\proj{\psi_{\max}} \right) S_{xn}^{F\dag}.
\end{align}
Note that the operation elements $S_{xn}^F$s are superposition-free and also satisfy $\sum_{x,n}S_{xn}^{F\dag}  S_{xn}^F= \mathbb{I}$.

\section{Implementation of super-operations using maximally-superposed operations \label{App:ImpSupOp}}
Here we outline how an arbitrary quantum super-operation can be implemented using maximally superposed operation, when consumed as a resource, and superposition-free operations. Consider an operation $\Phi$ and the maximally superposed operation $U_{\max}$, then what we desire to show is that
\begin{align}
 \tilde{\Omega}^F \left( \Phi \otimes U_{\max}  \right) \ \longrightarrow \ \tilde{\Omega} (\Phi).
\end{align}
In the first step, we bring the operations $\Phi$ and $U_{\max}$ together which are to operate on two different Hilbert spaces. In the second step, we apply a global superposition-free operation $\tilde{\Omega}^F$ and then trace out the second Hilbert space, which result in an arbitrary super-operation operating on $\Phi$. At the level of Choi matrix, the above implementation becomes
\begin{align}\label{eq:impl-SupOp}
 \Omega^F \left(C_{\Phi} \otimes \proj{\psi_{\max}}  \right) \ \longrightarrow \ \Omega (C_{\Phi}),
\end{align}
where $C_{\Phi}$ and $\proj{\psi_{\max}}$ are the Choi matrices correspond to the operations $\Phi$ and $U_{\max}$ (shown in Eqs.~\eqref{eq:Umax} and \eqref{eq:UmaxChoi}).
We shall adhere to Choi matrix based representation in the following derivations. Indeed, $\Omega (C_{\Phi})$ has to satisfy the condition given in Eq.~\eqref{eq:cond-supop}.

Let us first consider implementation of unitary super-operation $\Omega=U$, where $U=\sum_{i,j=0}^{d^2-1} U_{ij} \ketbra{\phi_i}{\phi_j}$. Following \cite{Chitambar16a, Dana17}, we can immediately find superposition-free super-operation elements, for $\alpha=0, 1, \ldots, d^2-1$, 
\begin{align}
S^F_{\alpha}= \sum_{i,j=0}^{d^2-1} U_{ij} \ketbra{\phi_i}{\phi_j} \otimes \ketbra{\phi_{\alpha}}{\phi_{(i+\alpha \mod d)}},
\end{align}
that are acting on the joint-space and satisfy $\sum_{\alpha=0}^{d^2-1} S^{F \dag}_{\alpha}S^F_{\alpha}=\mathbb{I}$. Application of these operation elements results in the implementation of the unitary super-operation, as in Eq.~\eqref{eq:impl-SupOp}, i.e.,
\begin{align}
 \sum_{\alpha=0}^{d^2-1} S^F_{\alpha} \left(C_{\Phi} \otimes \proj{\psi_{\max}}  \right) S^{F \dag}_{\alpha} \ \longrightarrow \ U (C_{\Phi}) U^\dag.
\end{align}

A more general quantum super-operation can be implemented by means of superposition-free super-operation and access to maximally superposed evolution, too. At the level of Choi matrix, consider a super-operation $\Omega$ that is operating in a $d$-dimensional Hilbert space on a Choi matrix $C_{\Phi}^S$, i.e., $\Omega(C_{\Phi}^S) = \sum_m E_m C_{\Phi}^S E_m^\dag$, with the super-operation elements $\{E_m \}$. This super-operation can be implemented following the protocol given in \cite{Dana17}. First, we attach an operation with equal resource of $C_{\max}$, i.e.
\begin{align}
 C_{\Phi}^S \longrightarrow C_{\Phi}^S \otimes \rho^{AB}_d,
\end{align}
where $\rho^{AB}_d= \proj{\psi_d}$, with the maximally entangled state $\ket{\psi_d}=\frac{1}{d} \sum_i \ket{\phi_i}^A \otimes \ket{\phi_i}^B \in \mathcal{H}_A \otimes \mathcal{H}_B $. The $\rho^{AB}_d$ can be deterministically created from $C_{\max}$ in $\mathcal{H}_A$ and $\proj{\phi_0}$ in $\mathcal{H}_B$, and then a (superposition-free) CNOT operation over $\mathcal{H}_{AB}$. Now, there is a set of superposition-free super-operations
\begin{align}
 L_{jkm}:= [\bra{\psi_d^{(jk)}}E_m \otimes \mathbb{I}]^{SA} \otimes U_{(jk)}^B,
\end{align}
where $\ket{\psi_d^{(jk)}}= \left( \mathbb{I}^A \otimes U_{(jk)}^B\right)\ket{\psi_d} $ and $U_{(jk)}=Z^jX^k$, such that
\begin{align}
 \tr_{AB} \left[ \sum_{jkm} L_{jkm} (C_{\Phi}^S \otimes \rho^{AB}_d)  L_{jkm}^{\dag}\right]  = \sum_m E_m C_{\Phi}^S E_m^\dag=\Omega(C_{\Phi}^S). \nonumber
\end{align}
Note $\sum_{jkm} L_{jkm}^{\dag}  L_{jkm} =\mathbb{I} $, and the definitions of $Z$ and $X$ are given in Eqs.~\ref{eq:Z} and \ref{eq:X} respectively.

\section{Local vs non-local superpositions \label{App:corr}}
Superposition is a basis dependent quality. For a set of reference bases, a quantum evolution with non-vanishing superposition can be made superposition-free or of different amount of superposition, by carefully choosing another set reference bases. This set of bases does not necessarily have to be orthonormal unitary ones. They could be a set of non-unitary orthonormal bases. In all these cases, the formalism presented above to quantify superposition can be extended.   

So far, we have considered quantum systems or operations as a whole. Here we turn to study superposition in the situation where the quantum operation is acted on bipartite quantum systems. Such an operation, in general, can be expressed as   
\begin{align}
 \Phi^{AB}(\rho_{AB}) = \sum_k E_k^{AB} \rho_{AB} E_k^{AB \dag},
\end{align}
where the operation $\Phi^{AB}$ is acting on the Hilbert space $\mathcal{H}_{AB}=\mathcal{H}_{A} \otimes \mathcal{H}_{B}$, corresponds to the parties $A$ and $B$.  The individual operation element can be expressed in terms of local operation bases ($\{ F_i^A \otimes F_j^B \}$), as 
\begin{align}
 E_k^{AB}=\sum_{ij} c_{kij} F_i^A \otimes F_j^B.
\end{align}
To quantify superposition, one could consider both local ($\{ F_i^A \otimes F_j^B \}$) as well as global operation bases ($\{ F^{AB}_k \neq F_i^A \otimes F_j^B \}$). For any operation, there could be many choices of local bases that give rise to same operations and they are all unitarily connected. 

The superpositions in the global bases are equivalent to the one discussed before, and,  in general, cannot encode any quality that is appearing essentially due to the presence of substructures of the Hilbert space. On the contrary, by using local bases, one could, in principle, differentiate between superpositions that have local and non-local contributions. Here we crudely classify operations into two classes; one with vanishing non-local superposition and other with non-vanishing non-local superposition. 

Global operations with \emph{vanishing non-local superposition} are the ones that can be made superposition-free by locally rotating the local operation bases. On the contrary, a global operation with \emph{non-vanishing non-local superposition} are the ones that have a non-zero superposition for all possible choices of orthonormal local operations bases. 

In the following, we go on to prescribe how to classify 
global operations in this regard. Furthermore, we quantify non-local superposition present in an operation $\Phi^{AB}$ by the amount of ``quantum'' correlation it generates in the state
\begin{align}\label{eq:corr}
\gamma_{\Phi^{AA^\prime BB^\prime}}= [ \Phi^{AB} \otimes \mathbb{I}^{A^\prime B^\prime}] \left( \proj{\psi}_{AA^\prime} \otimes \proj{\psi}_{BB^\prime}  \right), 
\end{align}
across the partition $AA^\prime $ and $ BB^\prime$. Here we denote the maximally entangled states as $\ket{\psi}_{AA^\prime}=\frac{1}{\sqrt{d_A}} \sum_{i=0}^{d_A-1} \ket{i_A i_{A^\prime}}$ and  $\ket{\psi}_{BB^\prime}=\frac{1}{\sqrt{d_B}} \sum_{i=0}^{d_B-1} \ket{i_B i_{B^\prime}}$. Depending on $\Phi^{AB}$, the state $\gamma_{\Phi^{AA^\prime B B^\prime}}$ can be uncorrelated or correlated across the partitions $AA^\prime $ and $ BB^\prime$. Moreover, the correlation present in the state could be classical, (separable) quantum correlations or entanglement and that can then be quantified by means of traditionally used measures introduced for quantum states. We shall not elaborate on these measures, as it is not absolutely necessary for our considerations here. However, an interested reader may consult with \cite{Horodecki09, Modi12}, for example.

\

\noindent {\it Operations with vanishing non-local superposition -- } Joint quantum operations that have vanishing non-local superposition are (L1) \emph{uncorrelated} operations 
\begin{align}
\Phi^{AB}_u (\rho_{AB})= \Phi^A \otimes \Phi^B (\rho_{AB}), 
\end{align}
and (L2) \emph{classical}-like operations 
\begin{align}
\Phi^{AB}_c(\rho_{AB})=\sum_i f_i^{AB} F_i^A \otimes F_i^B \rho_{AB} F_i^{A\dag} \otimes F_i^{B\dag}, 
\end{align}
where $\tr \left( F_i^{A/B} F_j^{A/B \dag} \right) = \delta_{ij} d_{A/B}$ for all $i,\ j$. 
It can be easily seen that the states $\gamma_{\Phi^{AA^\prime B B^\prime}}$, in Eq.~\eqref{eq:corr}, correspond to these operations are either uncorrelated or classically correlated state across the partition $AA\prime$ and $BB^\prime$, which means they could be made diagonal in orthonormal product bases. 

\

\noindent {\it Operations with non-local superposition -- } Quantum operations that possess non-zero non-local superposition ($\Phi^{AB} \notin \{ \Phi^{AB}_u, \Phi^{AB}_c  \}$) belong to the following classes.

\noindent (G1) \emph{Classical-quantum} like operations  
\begin{align}
\Phi^{AB}_{cq}(\rho_{AB})=\sum_{i} f_i^{A} F_i^{A} \otimes B_i \rho_{AB} F_i^{A\dag} \otimes B_i^\dag, 
\end{align}
where $\tr \left( F_i^{A} F_j^{A \dag} \right) = \delta_{ij} d_{A},  \ \forall i,\ j$ and $\exists i\neq j $ such that $\tr \left( B_i B_j^\dag \right) \neq 0$. 

\noindent (G2) \emph{Quantum-classical} like operations 
\begin{align}
\Phi^{AB}_{qc}(\rho_{AB})=\sum_{i} f_i^{B} A_i \otimes F_i^B \rho_{AB} A_i^\dag \otimes F_i^{B\dag}, 
\end{align}
where $\tr \left( F_i^{B} F_j^{B \dag} \right) = \delta_{ij} d_{B},  \ \forall i,\ j$ and $\exists i\neq j $ such that $\tr \left( A_i A_j^\dag \right) \neq 0$.

\noindent (G3) \emph{quantum-quantum} like operations 
\begin{align}
 \Phi_{AB}^q(\rho_{AB})=\sum_{i} A_i \otimes B_i \rho_{AB} A_i^\dag \otimes B_i^\dag,
\end{align}
where $\exists i\neq j $ and $\exists k\neq l $ for which $\tr \left( A_i A_j^\dag \right) \neq 0$ and $\tr \left( B_k B_l^\dag \right) \neq 0$ respectively.

\noindent (G4) \emph{Entanglement} like quantum operations that do not belong to any of the above classes, i.e., 
\begin{align}
\Phi^{AB}_e \notin \{ \Phi^{AB}_{u}, \Phi^{AB}_{c}, \Phi^{AB}_{cq}, \Phi^{AB}_{qc}, \Phi^{AB}_{qq}  \}. 
\end{align}

The non-local superposition can be quantified in terms of the quantum correlation in the state $\gamma_{\Phi^{AA^\prime B B^\prime}} $ across the partitions $AA^\prime$ and $BB^\prime$. Note that the operations belong to (G1), (G2) and (G3) will produce separable states with (discord like) classical-quantum, quantum-classical and quantum-quantum correlations respectively. In contrast, the operation in (G4) will produce a non-vanishing entanglement across the partitions $AA^\prime$ and $BB^\prime$.   

\section{Superposition, temporal order and causality \label{App:acausal}}
For any sort of quantumness to manifest, a certain form of superposition is necessary. As an example, for quantum correlation (e.g. quantum discord, entanglement, and non-locality) superposition in the bipartite product bases is necessary. However, the reverse is not true. In other words, the presence of superposition in product bases does not imply that there is entanglement or non-local correlations. In similar vein, it can be assumed that for any form quantumness in quantum evolutions to appear, superposition between evolutions is necessary. 
Here we consider four types of quantumness that exist in quantum evolutions; indefinite temporal order, temporal Bell correlation, quantum a-causality and indefinite causal orders in quantum evolutions, and examine the roles of superposition for such behaviors.

\

\noindent {\it Indefinite temporal order \cite{Araujo14, Procopio15} and temporal Bell correlation \cite{Zych17} -- } For any two quantum operations, $\Lambda_1$ and $\Lambda_2$, that are applied on a state $\rho_S$ of a quantum system $S$, there could be two different temporal orders. These are $\Lambda_{21}(\rho_S)=\Lambda_2 \circ \Lambda_1 (\rho_S)$, where $\Lambda_1$ applied before $\Lambda_2$, and $\Lambda_{12}(\rho_S)=\Lambda_1 \circ \Lambda_2 (\rho_S)$, where $\Lambda_2$ applied before $\Lambda_1$. For simplicity let us consider these operations to be unitary operations, denoted as $U_{12}=U_1 \circ U_2$ and $U_{21}=U_2 \circ U_1$ to be applied on pure state $\rho_S=\proj{\phi}_S$. Individually, they have definite orderings in their implementations. By using a control switch we can implement either a probabilistic mixture or a coherent superpositions of these operations with different temporal orders. Consider an additional control qubit in a state $\rho_c$ and controlled operation 
\begin{align}
U^{cS}= \proj{0}_c \otimes U_{12} + \proj{1}_c \otimes U_{21}.
\end{align}

Now depending on the choices of the initial state $\rho_c$ and the choice of measurements on the control bit after the appliance of $U^{cS}$, the effective reduced operations on the state $\rho_S$ would be different. For a control qubit $\ket{\varphi}_c=\sum_{i=0}^1 \sqrt{p_i} \ket{i}_c$, the joint operation on the control bit and system results 
\begin{align}
 \ket{\phi^\prime}_{cS}= U^{cS} [ \ket{\varphi}_c \otimes \ket{\phi}_S ].
\end{align}
Now the reduced dynamics on the system, by tracing out the control system, become a probabilistic mixture of ordered operations, i.e. $p_0 U_{12} \rho_S U^{\dag}_{12} + p_1 U_{21} \rho_S U^{\dag}_{21}$. By selectively projecting the control qubit in $\proj{0}$ ($\proj{1}$) the applied operation on the system can be controlled to $U_{12} \rho_S U^{\dag}_{12}$ ($U_{21} \rho_S U^{\dag}_{21} $).
Now instead, if we perform a selective projection $\proj{+}_c$ on the control bit and trace out, where $\ket{+}=1/\sqrt{2}(\ket{0}+\ket{1})$, then the effective reduced operation on the system becomes   
\begin{align}\label{eq:IndTempOrd}
 \left(\sqrt{\frac{p_0}{2x}} U_{12} + \sqrt{\frac{p_1}{2x}} U_{21}\right) \ket{\phi}_S,  
\end{align}
where the operation is implemented with a probability $x$. The resultant operation (in Eq.~\eqref{eq:IndTempOrd}) applied on the system does not have definite temporal order. It is composed of a coherent superposition of two evolution which are of different temporal orders. For $U_{12} \neq U_{21}$, the operation possesses a non-vanishing non-local superposition. In fact, for the situation where $U_{12}$ and $U_{21}$ orthogonal, the evolution can be collapsed to the one with a definite temporal order (i.e. to $U_{12}$ or $U_{21}$). It is clear from the definition that, the evolutions to have indefinite temporal orders, superposition between operations with definite orders are necessary.

Recently, Bell like temporal correlation has been proposed for bipartite quantum evolutions \cite{Zych17}. Consider a bipartite system that is composed of parties  $A$ and $B$. Each party can go through an evolution as a result of two operations applied one after another. Say for party $A$, the evolution is represented by $\Phi_{21}^A(\rho_A)=\Phi^A_2 \circ \Phi^A_1 (\rho_A)$, where the operation $\Phi^A_1$ is applied before $\Phi^A_2$. The order of appliance of the operations can be temporally reversed to give rise to another evolution $\Phi_{12}^A(\rho_A)=\Phi^A_1 \circ \Phi^A_2 (\rho_A)$, where the operation $\Phi^A_2$ is applied before $\Phi^A_1$. Similarly, the party $B$ can also go through two different evolution due to two temporal orders, say $\Phi_{21}^B(\rho_B)=\Phi^B_2 \circ \Phi^B_1 (\rho_B)$ and $\Phi_{12}^B(\rho_B)=\Phi^B_1 \circ \Phi^B_2 (\rho_B)$. Now there could be a control system which has access to both parties $A$ and $B$. Depending on control state and with a selective measurement, the control qubit can implement operations $\Phi^A_{12} \otimes \Phi^B_{12} (\rho_A \otimes \rho_B)$ and $\Phi^A_{21} \otimes \Phi^B_{21} (\rho_A \otimes \rho_B)$ selectively, or any of their coherent superposition. For simplicity, suppose the pure case scenario, where the operations are unitary, i.e. $\Phi^{A/B}_{1/2}=U^{A/B}_{1/2}$ and the states are pure $\rho_{A/B}=\proj{\phi}_{A/B}$. Now considering a control bit in a state $\ket{\varphi}_c=\sum_{i=0}^1 \sqrt{p_i} \ket{i}_c$, evolved using a joint control-systems unitary 
\begin{align}
 U^{cAB}=\proj{0} \otimes U^A_{12} \otimes U^B_{12} + \proj{1} \otimes U^A_{21} \otimes U^B_{21},
\end{align}
and then performing a selective measurement with the projector $\proj{+}_c$ on the control bit, we have
\begin{align}\label{eq:TempBell}
 \left(\sqrt{\frac{p_0}{2x}} U^A_{12} \otimes U^B_{12} + \sqrt{\frac{p_1}{2x}} U^A_{21} \otimes U^B_{21}\right) \ket{\phi}_A \otimes \ket{\phi}_B,  
\end{align}
which is implemented with a probability $x$. In case, where the unitaries $U^A_{12}$ and $U^A_{21}$, as well as $U^B_{12}$ and $U^B_{21}$, are orthogonal, the resultant state in Eq.~\eqref{eq:TempBell} becomes an entangled state, which violates the temporal Bell inequality \cite{Zych17}. Note these operations possess non-local superpositions. In fact, this entanglement is necessary to violate temporal Bell inequality, which in turn implies that non-local superposition in the evolution is a prerequisite to exhibit temporal Bell correlations, at least in the pure case scenario.  

\

\noindent {\it A-causality \cite{Beckman01, Eggeling02, Piani06} and indefinite causal order \cite{Oreshkov12, Brukner14, Rubino17} -- } In classical mechanics, events are bounded to respect certain causal order. If the events are space-like separated, the events occur independently and are causally disconnected. If the events are time-like separated, then the event in past can influence the one in future. Therefore they are causally connected and there is a definite order between events in past to future. Events in the future cannot influence the event occurred in the past. 

However, in the quantum domain, there are situations where the future event could, in principle, influence the past. The generation of such quantum situation depends on the quantum operations. For quantum operations that generate quantum events, where the past can only influence the future one,  are called semi-causal \cite{Beckman01} and they are semi-localizable \cite{Eggeling02} too. Therefore the operations that violate semi-causality cannot be semi-localizable. We denote the latter operations as the \emph{a-causal} operations. Instead of answering the question of how to implement a-casual operations, we consider here if the superposition between evolutions is necessary to exhibit a-causality.

Consider two parties $A$ and $B$ and a joint operation $ \Phi_{AB}$. The operation is semi-causal, in the sense that $A$ can signal $B$, iff there exist two different states on $A$, $\ket{\phi}_A$ and $\ket{\phi^\prime}_A$, such that 
\begin{align}
\tr_A\left( \Phi_{AB}( \proj{\phi}_A \otimes \proj{\phi}_B \right) )\neq \tr_A\left( \Phi_{AB}( \proj{\phi^\prime}_A \otimes \proj{\phi}_B \right)),  \nonumber  
\end{align}
for an arbitrary state $\ket{\phi}_B$ \cite{Beckman01}. This, in turn, implies that by changing a state on $A$, the outcome on $B$ can be modified and therefore $A$ can signal $B$. 
Consider an operation $\Phi^{A\rightarrow  B}$, acting on two-qubit system, with the operation elements 
\begin{align}
 E^{AB}_1=\proj{0}^A \otimes U_0^B \ \ \mbox{and} \ \ E^{AB}_2=\proj{1}^A \otimes U_1^B, \nonumber
\end{align}
where the unitaries are orthonormal, $\tr(U_0^{B\dag} U_1^B)=0$. The operation is superposition-free.
Now with the initial choices of $A$ states, the effective operation on $B$ can be controlled. Consider $\proj{0}_A \otimes \rho_B$ and $\proj{1}_A \otimes \rho_B$, or $\proj{\phi}_A \otimes \rho_B$, the resultant operation of $B$ are $U_0^B \rho_B U_0^{B\dag}$, $U_1^B \rho_B U_1^{B\dag}$ and $|\braket{\phi|0}_A|^2 U_0^B \rho_B U_0^{B\dag} + |\braket{\phi|1}_A|^2 U_1^B \rho_B U_1^{B\dag}$ respectively. Hence, non-local superposition is not necessary for the quantum operation to have semi-causal behavior, where $A$ signals $B$.

An operation is a-causal if $A$ could signal $B$ and also $B$ could signal $A$. Necessary and sufficient conditions for an operation to be a-causal are given in \cite{Beckman01, Eggeling02, Piani06}.  For any pure (unitary) evolution these conditions can be further simplified. It has already been noted that a unitary exhibits a-causal behavior if and only if the unitary is not a product of local unitaries, i.e. $U_{AB} \neq U_A \otimes U_B$ \cite{Beckman01, Piani06}. The joint unitary operation has to have non-vanishing non-local superposition. In other words, if the unitary assumes the form $U_{AB}=\sum_k c_k F^A_k \otimes F^B_k$, where $ \tr(F^{A/B}_k F^{A/B}_l)=\delta_{kl} d_{A/B}$ and $|\{k\}|>1$, then it manifests an a-causal behavior. The reverse statement is also true, so long unitary operations are concerned. 

However, the situation is very different if one goes beyond unitary operations. We can even show that non-local superposition is not necessary for the operations to have a-causal behavior. Even a classical like operation could give rise to a-causality.  To see this, we can simply construct an a-causal operation $\Lambda^{A \leftrightarrow B}$, acting on two-qubit system, where the operation elements are 
\begin{align}
& E_1=\frac{1}{\sqrt{2}}\proj{0}^A \otimes \sigma_x^B, \ \  E_2=\frac{1}{\sqrt{2}}\proj{1}^A \otimes \sigma_y^B, \nonumber \\
& E_3=\frac{1}{\sqrt{2}} \sigma_x^A \otimes \proj{0}^B, \ \ E_4=\frac{1}{\sqrt{2}}\sigma_y^A \otimes \proj{1}^B. \nonumber
\end{align}
It is easy to see that with this operation both $A$ and $B$ are able to signal each other. However, it does not have non-local superposition, as all local bases are mutually orthogonal to each other. Thus this operation exhibit an a-causal behavior and that is, even, without possessing non-local superposition. This operation indicates the fact that, to exhibit a-causal behavior, non-local superposition is not necessary.

Unlike a-causality, the study of \emph{indefinite causal order} relies on joint probability distributions, that arise due to an evolution of a system and a followed by measurement processes. The underlying principle could be based on any theory except for the fact that it has to respect definite local causal orders. In general, process matrices \cite{Oreshkov12} are used to characterize indefinite causal order. Entire formalism relies on probabilities and, in particular, certain inequalities in terms of linear combinations of these probabilities. That is why it is in general not straightforward to connect with superposition in the evolutions and the measurement processes. Furthermore, it is difficult to characterize quantum operations that exhibit indefinite causal order \cite{Araujo2017}. So far, the experimental demonstrations of indefinite causal order \cite{Rubino17, Goswami18} rely on quantum switches \cite{Chiribella13} and superposition of causal orders \cite{Chiribella12}. All these operations indicate that non-local superposition is necessary to exhibit indefinite causal order. However, an extensive study is required before it could be concluded unambiguously.

\end{document}